\documentclass{aa}
\usepackage{graphicx,subfigure}

\def\es{erg~s$^{-1}$}

\def\xmm{{\it XMM-Newton\/}}
\def\etal{et al.\ }
\def\betamod{$\beta$-model}

\def\kT {{\rm k}T}

\def\keV {\rm keV}

\def \rv {r_{200}}

\def \h50 {h$_{50}$}

\begin{document}

\title{An \xmm ~observation of the dynamically active binary cluster A1750}

\author{E. Belsole\inst{1,2} \and G.W. Pratt\inst{1} \and J-L. Sauvageot\inst{1} \and H. Bourdin\inst{3}}
\offprints{E. Belsole email:E.Belsole@bristol.ac.uk}

\institute{$^1$Service d'Astrophysique, CEA Saclay, L'Orme des Merisiers 
B\^at 709., F-91191 Gif-sur-Yvette Cedex, France\\
$^2$H.H. Wills Physics Laboratory, University of Bristol, Tyndall Avenue, Bristol BS8 1TL, U.K.\\
$^3$Observatoire de la C\^ote d'Azur, BP 4229, F-06304 Nice Cedex 4, France \\
\email{e.belsole@bristol.ac.uk}\\
\email{gwp@discovery.saclay.cea.fr}
\email{jsauvageot@cea.fr}
\email{bourdin@obs-nice.fr}
}

\date{Received ; accepted}

\abstract{We present results from the \xmm\ observation of the binary cluster A1750 at $z = 0.086$. We have performed a detailed study of the surface brightness, temperature and entropy distribution and confirm that the two main clusters of the system (A1750 N and A1750 C) have just started to interact. From the temperature distribution, we calculate that they are likely to merge sometime in the next 1 Gyr.  The more massive cluster, A1750 C, displays a more complicated temperature structure than expected. We detect a hot region associated with a density jump $\sim$ 450 kpc east of the cluster centre, which
appears to be a shock wave. This shock is not connected with the binary
merger, but it is {\em intrinsic} to A1750 C itself. From simple physical
arguments and comparison with numerical simulations, we argue that this
shock is related to a merging event that A1750 C has suffered in the past
1-2 Gyr. The larger scale structure around A1750 suggests that the system
belongs to a rich supercluster, which would presumably increase the
likelihood of merger events. These new \xmm\ data thus show that A1750 is a complex system, where two
clusters are starting to interact before having re-established equilibrium
after a previous merger. This merger within a merger indicates that  the present day morphology of clusters may depend not
only on on-going interactions or the last major merging event, but also on the
more ancient merger history, especially in dense environments.
}
\maketitle
%
\section{Introduction}
 Clusters of galaxies are the largest gravitationally bound structures in the Universe, and because of their observable properties and  sheer number, they represent invaluable cosmological probes.

The hierarchical scenario of structure formation in the Universe predicts the  growth of clusters of galaxies through accretion of smaller units by gravitational infall and mergers. During a merger event more than 10$^{63}$ ergs are dissipated in the intra-cluster medium (ICM) by shock heating, leading to strong variations of physical characteristics such as the temperature, density and entropy. Numerical N-body simulations have shown that mergers produce substructure which is detected in both the density and temperature distributions (Schindler \& M\"uller~\cite{schindlermuller93}; Roettiger et al~\cite{roettiger97}; Ricker~\cite{ricker98}). The temperature substructures survive for $\sim 4$--6 times longer than density substructures, thus the temperature distribution is a strong indicator of the cluster history and dynamical state. X-ray spectro-imaging observations are well suited to investigate in a deeper detail the observable characteristics of the merger event.  

Density substructure in X-ray observations of some low-redshift clusters (Forman \& Jones~\cite{formanjones82}; Mohr, Fabricant \& Geller~\cite{MFG93}), together with the detection of other substructure in optical observations (Escalera et al. \cite{escalera}), suggests that these clusters are dynamically young (Buote \& Tsai \cite{BuTs})

The first quantitative X-ray temperature maps were obtained with ASCA (e.g., Markevitch et al~\cite{marketal98}) but, as with the later BeppoSax observations (e.g., De Grandi \& Molendi~\cite{DeGM99}), resolution was low due to the energy dependent Point Spread Function (PSF) of these telescopes. Thanks to the high effective area and spatial resolution now available from \xmm\ and Chandra, precise spatially resolved temperature maps are now possible. Results confirm the increase in temperature in the merging regions (e.g., Arnaud et al. \cite{arnaud01a}; Markevitch \& Vihklinin \cite{MarkVihk01}; Markevitch et al.~\cite{mark02}; Neumann et al.~\cite{neumann03}) but the new observations have also detected new phenomena, such as the cold fronts (Markevitch et al. \cite{mark00}).

Substructures in clusters contain a fossil record of the merger history. Statistical studies of  cluster morphology can thus provide an important test of cosmological models of structure formation.  In practise, attempts to compare the prediction of numerical simulations of structure formation to quantitative statistical studies of nearby cluster morphology have been hampered by our current poor knowledge of the effect of cluster growth on morphology. Despite the large amount of data available, the dynamical evolution of the ICM and the relation between the galaxies and gas during a merger event is still poorly understood (see Buote \cite{buote02} for a review).  A better understanding of the physics of merger events, in particular the relaxation time of substructures (Nakamura \cite{nakamura95}) is required.

Hence a systematic study of merging clusters of galaxies would help a great deal in deepening our understanding of the process of cluster formation and evolution. Our \xmm\ Guaranteed  Time (GT) time program was established for this purpose. We selected a small sample of clusters displaying the signature of substructure (on the basis of ROSAT imaging analysis), and which could be considered as being in different {\em epochs} of a merger event. 

In this paper we present the first object in our sample: the binary cluster A1750. In optical, A1750 shows a multi-peaked galaxy distribution (Beers et al. ~\cite{beersetal91}; Ramirez \& Quintana ~\cite{RQ90}; Donnelly et al.~\cite{donnelly01}); the two major peaks have a radial velocity difference of $\sim 1300$ km s$^{-1}$. 
A1750 was observed in X-ray with {\em Einstein} (Forman et al. \cite{forman81}), and subsequently identified as a canonical binary cluster, though three peaks are clearly visible in the {\em Einstein} X-ray image. The cluster was later observed with ROSAT  and ASCA (Novicki et al.~\cite{Noveetal98}; Donnelly et al.~\cite{donnelly01}); three peaks are also visible in the $2^{\circ}$ ROSAT image. The latter authors combine ROSAT and ASCA observations, as well as the galaxy distribution, giving the first kinematic and dynamical description of the cluster. They detected an enhancement of order $30\%$ in the temperature between the two main peaks in the X-ray emission, and suggested that a shock region is developing in the gas by compression. However, Donnelly et al.~(\cite{donnelly01}) concluded that the two clusters could effectively be considered as isolated objects, as the hot gas in the interacting region did not have a strong impact in the global temperature estimate, and its contribution to the total density was weak. Donnelly et al.'s mass analysis indicated a mass ratio of order 1.3, but their dynamical analysis was not conclusive as to whether or not the system is bound.

In this paper we use the high sensitivity and spatial resolution of \xmm\ to gain new insights into this system. Taking advantage of the large field of view of \xmm, we observed the two main subclusters in a single pointing. The exceptional sensitivity of \xmm\ allows us to produce temperature and entropy maps with an accuracy which considerably surpasses previous attempts. We detect new features in all maps and relate these to the dynamical state of the system, arguing that, while the clusters are just beginning to interact, at least one of them (A1750 C), and possibly both, may not yet have relaxed from  a previous merger event.

Throughout the paper we assume $H_0 = 50$ km s$^{-1}$ Mpc$^{-1}$, q=0.5. In this cosmology 1 arcmin corresponds to 130 kpc at the mean redshift of the cluster ($z=0.086$). 


\section{Observations and data preparation}

\subsection{Observations}
A1750 was observed for 34 ks in July 2001  (Revolution 300) by \xmm. In this paper only data from the European Photon Imaging Camera (EPIC; Str\"uder et al. \cite{Struder}; Turner et al. ~\cite{Turner}) are considered. Calibrated event list files were provided by the \xmm\ SOC. The observations were obtained with the MEDIUM filter; the full frame mode was used for MOS and the extended full frame for pn. Throughout this analysis  single pixel events for the pn data (PATTERN 0) were selected, while for the MOS data sets the PATTERNs 0-12 were used.

The background estimates were obtained using a blank-sky observation consisting of several high-latitude pointings with sources removed (Lumb et al. ~\cite{lumbbkg}). The blank-sky background events were selected using the same selection criteria (such as PATTERN, FLAG, etc.) used for the observation events. Furthermore, the blank-sky background file was recast in order to have the same sky coordinates as A1750, ensuring that the source and background products come from the same region of the detector, reducing errors induced by any detector position dependence.

The source and background events were corrected for vignetting using the weighted vignetting method described in Arnaud et al. ~(\cite{Arnaudvign}). This allows us to use the on-axis response matrices and effective areas.

\begin{figure*}[]
\centering
\subfigure[] 
{
    \label{fig:fig1a}
}
\hspace{0cm}
\subfigure[] 
{
    \label{fig:fig1b}
}
\caption{(a) EPIC counts image in the energy band 0.3-7.0 keV; the image is not corrected for vignetting (b) Iso-intensity contours of the MOS1+MOS2 image in the energy band  [0.3-1.4] keV overlaid on the optical DSS image. Contours are logarithmic, and were obtained by generating contours after replacing  point sources with Poisson noise from surrounding annuli.}
\label{fig:fig1} 
\end{figure*}

\subsection{Background estimate}

The estimate of the background level is a crucial point since we are
interested in extended and low surface-brightness sources, were the background effects are important especially at large distance from the centre. 

The \xmm\ background consists of several components, which may be variable in time and in space distribution. The soft proton background is a time and flux variable component. Within some observations it is possible to find several periods of time where the mean flux level varies by a factor of order 10 or more.  To clean the data for this emission we reject all frames outside 2$\sigma$ of the mean value in the 10-12 keV (12-14 keV for pn) light curves for each camera, using the method of Pratt \& Arnaud~(\cite{PA02}) to determine the mean and $\sigma$ values. For this analysis, light curves were grouped in bins of 100 s. The observation was relatively clean: using these criteria  $\sim 10\%$ of the total exposure time was rejected. The final exposure times are 30.4 ks, 31.4 ks, 23.4 ks for MOS1, MOS2 and pn cameras respectively. The blank-sky background event files were then cleaned using the same criteria. 

A second component of the \xmm\ background is represented by the particle induced background, which dominates at high energy ($>$ 5 keV) and induces fluorescence lines (Al, Si, Cu, Au) from the shielding of the camera and the detector itself. This background component is effectively described by the blank-sky background, under the hypothesis that the variation of this particle background is small. It has been demonstrated that the particle background is variable at the $\sim 10\%$-level. This variation is sufficiently small that it can be taken into account by a normalisation in the [10-12] keV (MOS) and [12-14] keV band (pn) between the source  and the blank-sky background observations. The normalisation used in this work are 1.09, 1.11, 1.11 for MOS1, MOS2 and pn respectively.

Finally, we have to consider the astrophysical X-ray background, which is the combination of a soft (E$<$ 1.5 keV) component, mainly due to the local bubble, and a hard component  due to unresolved cosmological sources (mainly AGN). This emission is properly an X-ray component (i.e., not particle-induced, or due to soft proton flares). The hard component of this astrophysical X-ray background is well taken into account by the subtraction of the blank-sky background. At lower energies, however, the astrophysical X-ray background is variable across the sky (e.g., Snowden et al.~\cite{snowden}). To account for this, one possibility is to use a local background as described in Pratt, Arnaud, Aghanim~(\cite{paa01}) and Arnaud et al.~(\cite{arnaud02}). However, emission from A1750 fills the whole field of view (FoV) and there is no large region which can be considered free of cluster emission with enough confidence to be used as a local background. In this analysis, the blank-sky background represents the best estimate we can give of the background associated with the cluster. However, the examination of radial profiles in several different bands suggests that at large radii the emission remaining after the subtraction of the blank-sky background is essentially of order zero, giving us confidence in our use of the blank-sky background only.


\section{Morphology}

\subsection{Image}

Figure~\ref{fig:fig1a} shows the combined MOS1, MOS2 and pn image of A1750 in the energy band 0.3-7.0 keV. The image is in counts, has not been corrected for exposure, and the regions outside the field of view (FoV) have not been masked. A total of 356000 photons have been collected by EPIC in this energy band. The double distribution of the emission is clear, with one emission peak at the centre of the FoV and the other to the north-east. There are also many serendipitous  point sources and some extended sources: their positions and fluxes were obtained using the EMSRLI file in the pipeproducts, and all those with flux greater than 10$^{-14}$ ergs s$^{-1}$  arcmin$^{-2}$  were masked throughout this analysis.

Contours of the low-energy (0.3-1.4 keV), adaptively-smoothed EPIC/MOS image are shown overlaid on the DSS image in Fig.~\ref{fig:fig1b}. The X-ray emission peaks are at [$\alpha=13^h30^m49^s.881$, $\delta=-01^\circ51\arcmin46\farcs70$] and [$\alpha=13^h31^m10^s.941$, $\delta=-01^\circ43\arcmin41\farcs65$] for the cluster at the centre of the FoV (hereafter A1750 C) and the north-eastern cluster (hereafter A1750 N) respectively.

The isophotes at the centre of A1750 C are elongated and their ellipticity decreases with the distance from the centre.  The distance between the peak of X-ray emission of A1750 C and the cD galaxy visible in the DSS image is $\sim 15\arcsec$ (32.5 kpc). This shift is larger than the possible attitude error of \xmm . The cD galaxy is offset from the X-ray peak towards the east, in which direction there is a clear compression of the isophotes. 

 The centroid of the X-ray emission of A1750 N is located exactly over the superimposed optical emission of two central galaxies. The X-ray isophotes in the central regions are quite  circular but they show a slight compression to the east-south-east. The abrupt change in the shape and orientation of the isophotes at large scale is however remarkable, where they become very elliptical and oriented in the direction of the line joining the centres of the two clusters. 


\subsection{2D $\beta$ model fitting}\label{sec:model2D}

To take into account the binary nature of the cluster and to give a quantitative measure of the morphological characteristics we observe,  and in an attempt to detect density substructures, we performed a bi-dimensional analysis.

We adopted the hypothesis that a $\beta$ model (Cavaliere \& Fusco-Femiano \cite{CFF76}) is an appropriate description of a relaxed isothermal cluster. We fit the surface brightness distribution of A1750 at low energy (where the density distribution is least temperature dependent) with a $\beta$ model and quantify the deviation from this model and the presence of substructures.  In fact, the interest of this exercise is to compare A1750 with the surface brightness distribution of two relaxed clusters.
\begin{table}[]
\begin{center}
\caption{Best fit results of the 2D $\beta$ model.}
\label{tab:2Dbetamod}
\begin{tabular}{l|cc}
\hline
 		& A1750 N & A1750 C \\
\hline
$r_{c1}$ (kpc) 	& 250 & 230 \\
$r_{c2}$ (kpc) 	& 180 & 200 \\
$\beta$		& 0.56 & 0.68 \\
$\alpha_0$ (J2000)& $13^h31^m10^s5$ & $13^h30^m49^s0$ \\
$\delta_0$ (J2000) &$-01^\circ43\farcm20\farcs61$ & $-01^\circ51\farcm54\farcs14$ \\
PA 	& $-12^{\circ} 12\arcmin 53\arcsec$ & $24^\circ 38\arcmin 14\arcsec$ \\
\hline
\end{tabular}
\end{center}
\end{table}

\begin{figure}[]
\begin{centering}
\caption{EPIC/MOS adaptively smoothed image with, overlaid, contours of the residuals after the subtraction of a double $\beta$ model. The residuals are traced at 2,3,5 and 7 $\sigma$ above the background. Further details in the text.}
\label{fig:resid2D}
\end{centering}
\end{figure}

For the fit we followed the prescription of Neumann \& B\"ohringer (1997) and apply this analysis to the MOS camera only because the large gaps in the pn camera are relatively difficult to take into account. Images of pixel size of 4\farcs1 from the weighted  MOS event files and in the energy band 0.3-1.4 keV were created, and then summed. Since the weight correction is applied directly to the events, the statistical Poisson distribution of photons does not apply for these images. To correctly take into account the errors, the variance  is given by $\sigma^2=\Sigma_iw_i^2$ where w$_i$ is the weight associated to the $i$\,th photon (see Arnaud et al.~\cite{Arnaudvign}). The error images were generated for each MOS image and the quadratic sum of the latter gives the summed error image. The combined MOS image was smoothed with a Gauss filter of $\sigma=15\arcsec$, chosen to be of the same order of the MOS PSF. This smoothing allows us to be confident that in each spatial interval of integration (meta-pixel) we can use the Gaussian statistic and find the best $\chi^2$ when fitting. The error image was treated according to the error propagation function for Gaussian filtered images described in Neumann \& B\"ohringer (1997).

Because of the double nature of A1750, the 2D $\beta$ model for a single cluster (see Neumann \& B\"ohringer~\cite{NB97}; Pratt \& Arnaud~\cite{PA02}) has been modified in order to take into account simultaneously the two clusters. The background was approximated as a unique constant value. The region between the two clusters was masked before fitting so as to reduce the interference from a possible interaction region. Point sources were also masked. Results of the best fit double 2D $\beta$-model are shown in Table \ref{tab:2Dbetamod}. The best fit model was then subtracted from the image in order to quantify the significance of any possible excess flux. The significance was calculated following the Appendix of Neumann \& B\"ohringer (1997). The positive residuals, traced at 2, 3, 5 and 7 $\sigma$ above the background are shown in Fig.~\ref{fig:resid2D}. 

The image clearly shows significant residuals in the centre of each cluster. The core of A1750 N is off-centre with respect to the structure at larger scale (which is well described by the $\beta$-model). The same effect is marginally  visible at the centre of A1750 C. Other significant residuals, other than the two cores, are detected to the north of A1750 N and to the west of A1750 C.

 This analysis does not show any significant substructure in the region between the two clusters (hereafter the middle region), with the exception of two marginally detected ($\sim 2\sigma$) point sources and a further source which is detected at greater than $5\sigma$. From comparison with the raw image, it is clear that this source is extended.

We have considered the nature of the extended source in the middle region. It is clearly detected in emission in the pn image (see Fig.~\ref{fig:fig1a}), but unfortunately most of the emission is located in the gaps of the MOS images. We extracted the spectrum of this source in a circle of 45\arcsec. As a background we used an annular region centred on the source, with an inner radius 1\farcm3 and outer radius 2\farcm2. The spectrum does not contain  emission lines strong enough to estimate the source redshift. We fitted the spectrum with an absorbed MEKAL model with redshift and column density parameters fixed to those of A1750 ($z=0.086$ and $n_H = 2.37 \times 10^{20}$ cm$^{-2}$). The best fit gives k$T=4.6^{+2.3}_{-1.3}$ keV and a flux of $1.3 \times 10^{-13}$ erg cm$^{-2}$ s$^{-1}$ in the energy range [0.3-10.0] keV. 
A superposition of the residual contours on the DSS image does not show any obvious optical counterpart.


\section{Thermal structure}

The lack of significant substructure in the middle region seems to suggest that there is little or no interaction between the two components of the cluster (however, see Sect.~\ref{sec:current}). In other words, the merger event is sufficiently early that the density enhancement expected in the interaction region (e.g. Roettiger et al. 1997; Ritchie \& Thomas 2002) has not yet been produced (or is only just beginning). Knowledge of temperature structure is thus a further important step toward the comprehension of the dynamical state of the system.

The spectral analysis was performed by extracting spectra from the weighted source and background (blank-sky) event lists. The out of time events affecting the pn observation were not taken into account, but in our experience they are not expected to have a detectable effect on the results. The response files used in this analysis are: m1\_mediumv9q20t5r6\_all\_15.rsp (MOS1),  m2\_mediumv9q20t5r6\_all\_15.rsp (MOS2) and epn\_ef20\_sY9\_medium.rsp (pn).

\begin{figure}[h!]
\begin{centering}
\caption{EPIC/MOS adaptively smoothed image in the energy range 0.3-0.6 keV  with superimposed contours of the 3 regions of interest for the global spectrum analysis. The image is logarithmically scaled.}
\label{fig:3zones}
\end{centering}
\end{figure}

\subsection{Global spectrum}
\label{sec:globspec}

We obtained a global spectrum for each of the regions displayed in Fig. \ref{fig:3zones}. The regions were chosen to give a first measure of the spectral characteristics of the two clusters and the middle region, and also serve as a useful comparison with previous results. The particular shape of the middle extraction region is determined by the exclusion of point sources and cluster emission. The spectrum of the background was extracted in the same region and, after normalisation, subtracted from the source spectrum.

\begin{figure}[h!]
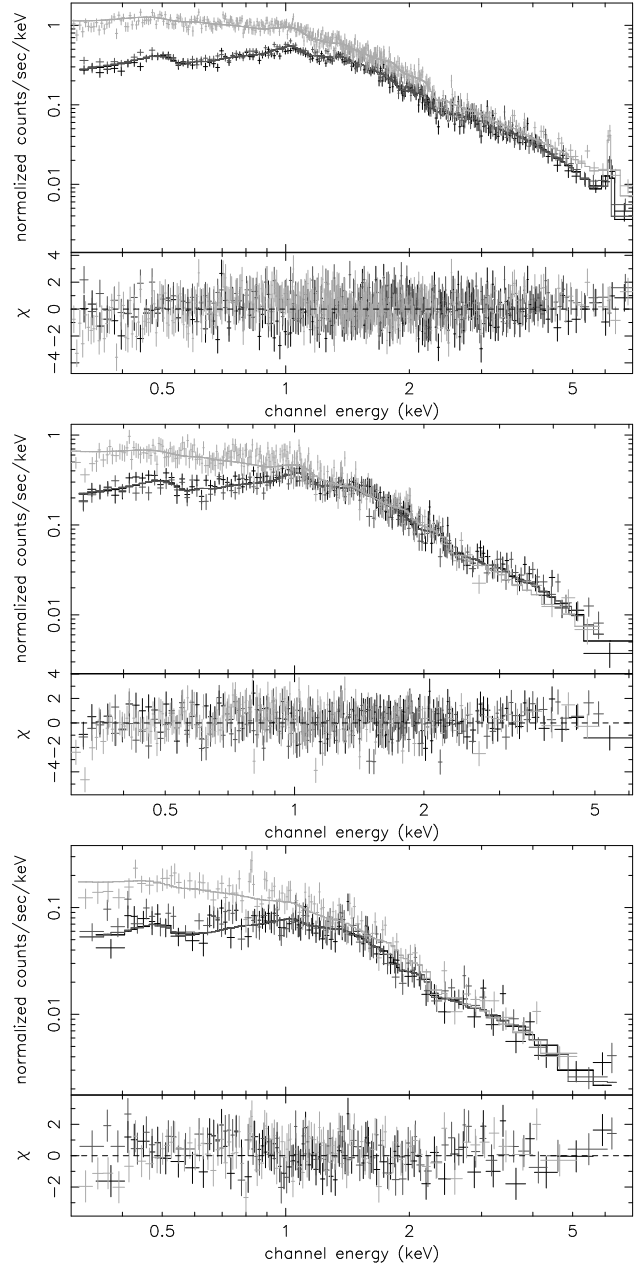

\centering
\includegraphics[scale=0.35,angle=-90,keepaspectratio]{f4a.ps}
\includegraphics[scale=0.35,angle=-90,keepaspectratio]{f4b.ps}
\includegraphics[scale=0.35,angle=-90,keepaspectratio]{f4c.ps}
\caption{Global spectra of the three regions in Fig.~\ref{fig:3zones}:A1750 C (Top); A1750 N (centre); middle region (bottom). MOS spectra are in black, the pn in grey. The three camera spectra were fitted simultaneously.}
\label{fig:spglob}
\end{figure}

\begin{table*}
\begin{center}
\caption{Results of the global spectrum analysis in the region shown in figure \ref{fig:3zones}. The column density is obtained by fitting the MOS spectra only and then fixed to this value when combining with the pn spectrum. The value of the  column density listed here is the best fit MOS value (see text for more details). Errors are at 90\% for one significant parameter.}
\label{tab:spglobfit}
\begin{tabular}{cccccc}
\hline
region	 & $N_H$ & kT    & $Z$             & $\chi^2$/d.o.f & $L_X$ [0.2-10] keV \\
		 & ( $10^{20}$ at/cm$^{-2}$) &(keV)    & solar             &  & (\es) \\

\hline
Cluster ~C & 2.18$\pm0.44$ & 3.87$\pm$0.10 & 0.32$\pm$0.04 & 1217.79/1139     & 2.2 10$^{44}$ 			\\
Cluster ~N & 1.00$\pm0.8$ & 2.84$\pm$0.12 & 0.22$\pm$0.05 & 797.33/663      & 1.3 10$^{44}$			\\
Middle  & 0.5($<1.4$) & 5.12$^{+0.77}_{-0.69}$ & 0.27$^{+0.28}_{-0.19}$ &  418.32/426 & 1.7 10$^{43}$			\\
\hline
\hline
\end{tabular}
\end{center}
\end{table*}

 The spectra of the 3 cameras, excluding the channels below 0.3 keV, were simultaneously fitted with an absorbed MEKAL model with the temperature, metallicity  and normalisation (emission measure) as free parameters.  
The column density was fixed to the best fit value obtained by fitting the MOS cameras only. This value is in agreement with the galactic value for A1750 C. (The spectrum of A1750 N and the middle region are better fitted with a lower than Galactic absorption). The best fit results are listed in Table \ref{tab:spglobfit}. The temperature in the middle region is significantly higher (by $\sim 30\%$) than that of A1750 C, in agreement with the results of Donnelly et al.~(\cite{donnelly01}). In order to be confident in this detection of a significantly higher temperature, we also investigated the dependence of the temperature of middle region spectrum on the Galactic absorption. If we fix the $N_H$ to the Galactic value, the temperature drops to  4.35$\pm 0.45$ keV (for  a reduced $\chi^2$=425.86/427 d.o.f.), while if we fix the $N_H$ to the value obtained by fitting cluster N, the best fit temperature is 4.9$^{+0.6}_{-0.5}$ keV ($\chi^2$=418.98/427 d.o.f.). In both cases the temperature is higher than the temperature of A1750 C at the $90\%$ confidence level, and the goodness of the fit is clearly in favour of a lower absorption. In Table \ref{tab:spglobfit} we list the temperatures obtained by fitting simultaneously the three cameras with the column density fixed  to the best fit MOS value in each region. The spectral fits are shown in Fig. \ref{fig:spglob}.

For A1750 C and the middle region, our results are in excellent agreement with the ASCA analysis (Donnelly et al. 2001), but we find a lower temperature than for A1750 N, if the mean temperature between zones 1 and 2 in Donnelly et al.'s paper is taken. However, if we compare our result with the ASCA temperature in region 1 of Donnelly et al.'s Figure 2, the two results are in better agreement.

We further fit the global spectrum of A1750 N with a two temperature model. In this case we find $kT_1=3.17\pm0.1$ keV, $kT_2=0.65\pm0.17$, with abundances of 0.3$\pm0.7$ solar and a $\chi^2$=754.05/660 d.o.f. According to the F-test, it is $99\%$ probable that the 2 temperature model is a better representation of the data, suggesting that the gas is multi-temperature (see also Sect.~\ref{sec:1danal}). 


\subsection{Temperature map}
\label{sec:tmap}
\subsection{Adaptive binning}\label{sec:tmapbox}

The high sensitivity of \xmm\ allows us to build a temperature map. To optimise the spatial resolution, without any spatial a priori, we have defined a specific binning which follows the statistics, and leads to fairly homogeneous errors in the temperature determination for each meta-pixel. This approach is similar to the one presented by Sanders \& Fabian~(\cite{sf01}). We produce a full energy band image with sources excised. The algorithm starts from the largest spatial scale and continues down to the smallest possible, dividing by 4 each cell of the image containing more than a given number of photons (here we choose 2000 photons for the 3 cameras in the full energy band prior to background subtraction). We end up with a list of cells that have more or less the same number of photons. We apply this spatial binning to the observation and to the background to extract all spectra. Then using XSPEC, we fit the MEKAL model (with the $N_H$ fixed to the galactic value and abundances set to 0.3 solar) to each spectrum. The fit temperature of each cell is then used to build the temperature map shown in Fig.~\ref{fig:tmap}. The typical error (which depends slightly on the temperature) is around $\pm 0.6$ keV.

As already suggested by the broad global spectrum analysis above, the region between the two clusters displays a higher temperature, implying a dynamical interaction. The map is in excellent agreement with that of Donnelly et al.~(\cite{donnelly01}), but because of the improved sensitivity, our map indicates that the temperature structure is more  complicated.  

A1750 N displays a relatively uniform temperature with a suggestion of a cooler core region (although the large bins are not optimised). To the west side of A1750 N we observe a higher temperature which may be connected with the interaction between the two clusters. A hot region to the south of the centre of A1750 N should be also noted, which interestingly lies ahead of the compression of the isophotes near the core. 

A new feature, observed for the first time in these data, is the hot region $\sim$3-4 arcmin to the east of the centre of A1750 C. This hot region corresponds exactly to a compression of the isophotes in the X-ray surface brightness, and does not appear to be related to the larger hot region due to the interaction between the clusters. In other words, it appears to be {\it intrinsic} to A1750 C itself.

\begin{figure*}[!]
\centering
\subfigure[] 
{
    \label{fig:tmap}
}
\hspace{0cm}
\subfigure[] 
{
    \label{fig:tmapwlet}
}
\hspace{0cm}
\subfigure[] 
{
   \label{fig:smap}
}
\hspace{0cm}
\subfigure[] 
{
    \label{fig:tmaperr}
}
\vspace{-0.5cm}
\caption{Temperature and entropy maps of A1750. Temperature colour scales are in keV. {\bf a}: Temperature map obtained by extracting spectra in adaptive bins as described in Sect. \ref{sec:tmapbox}.{\bf b}: Temperature map derived by applying the wavelet algorithm as described in Sect. \ref{sec:wlets}. This temperature map was obtained using data from the MOS cameras only. {\bf c}: Specific entropy map. Scale is in arbitrary units. {\bf d}: Colour coded map of the error in temperature. Pixels with errors greater that 1.5 keV have not been taken into account.}
\label{fig:wlet} 
\end{figure*}


\subsection{Wavelet temperature analysis}\label{sec:wlets}

We also computed a temperature map by applying the new
multi-scale spectro-imagery algorithm described in  Bourdin et al.~(\cite{bourdin03}). In that paper, the algorithm was tested on simulated
observations; here we present preliminary results from the first application to a real observation.
For a detailed description of the method see Bourdin et al.~(\cite{bourdin03}).

The A1750 temperature map was computed using only the event lists from the
MOS cameras, as we do not yet have a stable model for the pn background. Due to the steep decrease of the cluster emissivity with radius, the analysis has been restricted to regions where significant signal is detected at more than one scale. The
background normalisation was fixed to a value estimated for each camera
in an external region of the field of view, where the background
emissivity dominates.
The temperature map is a $128 \times 128$ pixels image, where the multi-scale analysis has been performed on 5 scales in order to prevent large scale edge effects, so that details on spatial scales from around 15\arcsec\ to 4\arcmin\ are expected. 
Spectral fits with error-bars
higher than 1.5 keV are not taken into account due to evidence for a
bias of the temperature estimator in regions with high background level
. The issue of background modelling for spectral fits in each element is the major source of uncertainty, especially at large distance from the centre where the background itself begins to dominate. To be conservative, we have applied {\em a posteriori} a mask to the wavelet maps defined such that at least 10 photons are detected in the emissivity wavelet map. This corresponds to a 3$\sigma$ signal above the background.

Since we have not yet perfected the removal of point sources in the
algorithm, the following wavelet-derived maps are presented
without point sources removed. The side by side comparison of the adaptive
binning and wavelet temperature maps (Figs 5a and 5b) allows the interested
reader to judge the effect of the point sources on the temperature structure.

Note that, in the wavelet map, all the point sources appear in the highest
frequency plane. This means that their contribution to the local
temperature is limited to that spatial frequency (in other words they
 always appear in the temperature map at the highest frequency, i.e., as
point-like sources).

In general, the wavelet-derived temperature (Fig.~5b; errors Fig.~5d) map is in good agreement with the adaptively-binned temperature map. We note the following:

\begin{itemize}
\item The hot rectangle to the South of A1750 C in the adaptively binned
map is not seen in the wavelet map, but this region is at the periphery of
the cluster emission and has large errors (k$T$=5.7$^{+1.9}_{-1.7}$ keV).

\item The hot regions appearing in the wavelet map to the south and
south-west of the centre of A1750 C appear to be due to point sources.

\item The hot region between the clusters is recovered in the wavelet
map, but is compromised by a point source. Because of this, we are unable
to determine if the structure is continuous or clumpy.

\item The temperature structure the the east of the centre of A1750 C
(discussed in more detail later on) is robust, as there are no associated
point sources. It is resolved in the wavelet map into two arc-like
structures of higher temperature. The first is located at $\sim 1\farcm8$
from the centre of A1750 C to the east. The second one starts at around 3\farcm5 in the same direction. These two hot regions are separated by a gap of lower temperature.

\item Finally, the hot region at the Northwest end of the arch between
the clusters appears to be stable, appearing in both maps, although some of this emission may be due to a strong point source.

\end{itemize}

The western side of A1750 C appears to be relatively isothermal. A1750 N displays a quite uniform temperature distribution, with an indication of a higher temperature in the centre than that observed with the adaptive binning map (but see also Section \ref{sec:1danal}).

Note that the wavelet temperature map has been derived using only the MOS data, while the adaptive binning map uses data from all three cameras. As such, we would expect the wavelet map to only have approximately half the sensitivity of the adaptively-binned map. This loss of sensitivity is crucial in  regions of low surface brightness, such as the middle region. Despite these caveats, the agreement between the maps is good.


\section{Entropy}

The multi-scale spectro-imagery algorithm described above can be
applied to estimate the projected spatial distribution of other
parameters of the intra-cluster gas, provided
that the expected parameter and its fluctuation are estimated within
each resolution element. In this paper, we applied the algorithm to the EPIC-MOS data
to compute a crude  map of the the gas specific entropy , which is convenient to define in terms of observed quantities  as $S \propto T/n^{2/3}$, as a function of the intra-cluster gas temperature $T$,
and density $n$\footnote{more precisely, the definition of ``entropy'' is the logarithm of the quantity above plus a constant (see Ponman et al. \cite{psf03} and references therein)}.

To do so the averaged cluster emissivity per surface unit ($N$) and
temperature ($T$) are fitted simultaneously within the different
resolution elements, following the same procedure as for computing the
temperature map. Since the cluster emissivity ($N$) scales as
to the square of the density of the intra-cluster gas ($n$), both
quantities leads to a crude estimator of the averaged specific entropy, $\hat{S} = \frac{T}{{N}^{1/3}}$.

Just like for the temperature map, the 'entropy'  map 
is obtained by
computing maps of $\hat{S} \pm \sigma_{\hat{S}}$  at five different scales,
leading to a wavelet transform. Then the wavelet coefficients are
thresholded in order to remove the noise contribution, again following a $1
\sigma$ significance criterion.

The entropy map, shown in Fig.~\ref{fig:smap} is very similar to entropy maps obtained in numerical simulations of cluster mergers (e.g. Ricker \& Sarazin \cite{RS01}). We observe a lower entropy in the centres of the two clusters, with an increase towards the external regions. The entropy distribution of A1750 N is very elongated towards the north, mainly following the emissivity distribution. We observe that the eastern side of A1750 C displays an enhancement in the entropy which does not follow a spherical distribution. The projected entropy distribution is quite turbulent in this region. The entropy in the middle region is also higher than in the centre of the two clusters, as expected from shock heating.

\section{One dimensional analysis}\label{sec:1danal}

The two dimensional analysis above has revealed a wealth of new detail in both the morphology and temperature of the gas in this cluster. In this Section we use the two dimensional results to divide each subcluster into discrete regions, with the aim of undertaking a classical 1D  analysis for comparison with previous results. The regions we have chosen are shown in Figure~\ref{fig:1dregions}. These regions are:

\begin{figure}[]
\begin{centering}
\caption{Adaptively-smoothed MOS image in the energy band 0.3-6.0 keV. The lines limit the regions used to extract 1 dimensional profiles. The image in in logarithmic scale.}
\label{fig:1dregions}
\end{centering}
\end{figure}

\begin{itemize}
\item Region 1 (R1) corresponds to the part of A1750 N which is relaxed-looking and fairly isothermal. R1 has been defined by excluding all data between position angles $77^\circ - 330^\circ$ (anticlockwise from N, centred on $\alpha = 13^h31^m00^s; \delta = -01^{\circ}46\arcmin46\arcsec$ ), related to all hot regions connected with the interaction.

\item Region 2 (R2) is that part of A1750 C which is relaxed-looking and very isothermal. R2 is defined by a sector between position angles $205^\circ - 354^\circ$, anticlockwise, centred on $\alpha=13^h30^m49^s.881$, $\delta=-01^\circ51\arcmin46\farcs70$, the A1750 C emission peak.

\item Finally, Region 3 (R3) is delimited by a sector between position angles $65^\circ - 205^\circ$ (anticlockwise), centred on the A1750 C emission peak. R3 includes the zone  $\sim 4\arcmin$ the east of the centre which shows the hot, arc-like temperature feature and isophotal compression.

\end{itemize}

\subsection{Data analysis}

\subsubsection{Surface brightness profiles}

In each Region, we extracted azimuthally averaged surface brightness profiles from the observation and the corresponding Region in the blank-sky background. Photons were binned in circular sector annuli of width $5\arcsec$ centred on the peak of the emission. We use the energy band [0.3 - 1.4] keV to minimise the dependence of the emissivity on the temperature; the upper energy limit was set to avoid the instrumental Al and Si lines\footnote{We checked the emissivity-temperature dependence of the FeL blend by extracting profiles with the [0.9 - 1.2] keV band excluded. The best fitting \betamod\ parameters did not change significantly in this case, thus to maximise the S/N, we use the entire [0.3 - 1.4] keV band.}.

The background surface brightness profiles were subtracted from each corresponding cluster profile, using the appropriate [10. - 12.]([12. - 14.]) MOS(pn) keV normalisation factor. Each camera was treated separately. In all cases, the MOS and pn profiles agree well barring the expected normalisation differences, and so in each case the MOS and pn profiles were co-added.

Except in one case (R3), the surface brightness profile was binned such that (i) at least a $S/N$ ratio of $3\sigma$ was reached, and (ii) the width of the bins increased with radius, with $\Delta(\theta) > 0.1 \theta$. The logarithmic binning ensures a roughly constant $S/N$ in the outer parts of the profile. 


\subsubsection{Temperature profiles}

Radial temperature profiles were produced for each Region using the same mask as was used for the corresponding surface brightness profiles. Spectra were extracted in circular sector annuli centred on the peak of the emission. The widths of the annuli were chosen so that a minimum $5\sigma$ detection was achieved in the [2.0 - 5.0] keV band. A minimum width of $30\arcsec$ was also imposed, corresponding to the $90\%$ encircled energy radius of the MOS PSF. The blank-sky background spectra were extracted from the same regions and subtracted fom the source spectra. The background subtracted spectra were binned to $3\sigma$ to allow the use of Gaussian statistics.

Spectra from the three cameras were fitted simultaneously with an absorbed MEKAL model, with the column density fixed to the best-fit of the corresponding global spectrum (Tab.~\ref{tab:spglobfit}). The abundance for the fits was frozen at the best-fit global value for whichever cluster was under consideration\footnote{We did fit the radial profiles with spectra where the abundance was a free parameter: at $90\%$ confidence level however, the temperature profiles do not show significant differences from those obtained with the abundance frozen at the global value.}. The MOS spectra were fitted in the range [0.3-10.0] keV, and the pn spectra in the range [0.4-10.0] keV. 

We stress that we have not attempted to undertake a full PSF and deprojection analysis and that the results presented here are {\it projected} temperature profiles. PSF effects will be important if the cluster possesses very significant temperature gradients, or if the surface brightness profile is very peaked. We have taken care to minimise PSF effects by ensuring that the bins have a minimum width greater than the $90\%$ encircled energy radius of the PSF, and we have already seen that the temperature gradients, while significant, are relatively mild (not nearly so strong as in a cooling flow, for example).

The temperature plots below are presented with the upper X-axis in arc-minutes. The lower X-axis shows the radius in terms of the virial radius, calculated using the average temperature from the global spectral fit (Sect.~\ref{sec:globspec}), and the $r_{200}-T$ relation of Evrard, Metzler \& Navarro~(\cite{emn96}) at the cluster redshift, viz:

\begin{equation}
\label{eq:emn96}
r_{200} = 3690~(T/ 10~{\rm keV})^{1/2} (1+z)^{-3/2}.
\end{equation}


\subsubsection{Gas density profiles}

The surface brightness profile of each Region was fitted with various parametric models, all of which were convolved with the \xmm \ PSF (Ghizzardi 2001, Griffiths \& Saxton 2002) and binned in the same way as the observed profile. The surface brightness profile, at low energy, is generally well represented  by a \betamod:
\begin{equation}
$$S(\theta) = S_0 \left(1+\frac{\theta^2}{\theta^2_c}\right)^{-3\beta + 0.5} $$
\label{eq:sbprof}
\end{equation}

where $S_0$ is the central intensity, $\theta_c$ the core radius and $\beta$ the slope. In the hypothesis of an isothermal cluster, the two parameters $\theta_c$ and $\beta$ are related to the gas density profile by:
\begin{equation}
$$ n_{gas}(r) = n_0 \left(1+\frac{r^2}{r^2_c}\right)^{-3\beta/2}$$.
\label{eq:neprof}
\end{equation}

In addition to the standard \betamod\ above, we have also used the BB parametric model (Pratt \& Arnaud~\cite{PA02}), which is a double isothermal \betamod\ which assumes that both inner and outer gas density profiles can each be described by a \betamod, but with different parameters. The boundary between the two regions is a free parameter of the fit, and the density distribution, and its gradient, are continuous across the boundary.

\begin{figure}[!]
\begin{centering}
\includegraphics[scale=1.,angle=0,keepaspectratio,width=\columnwidth]{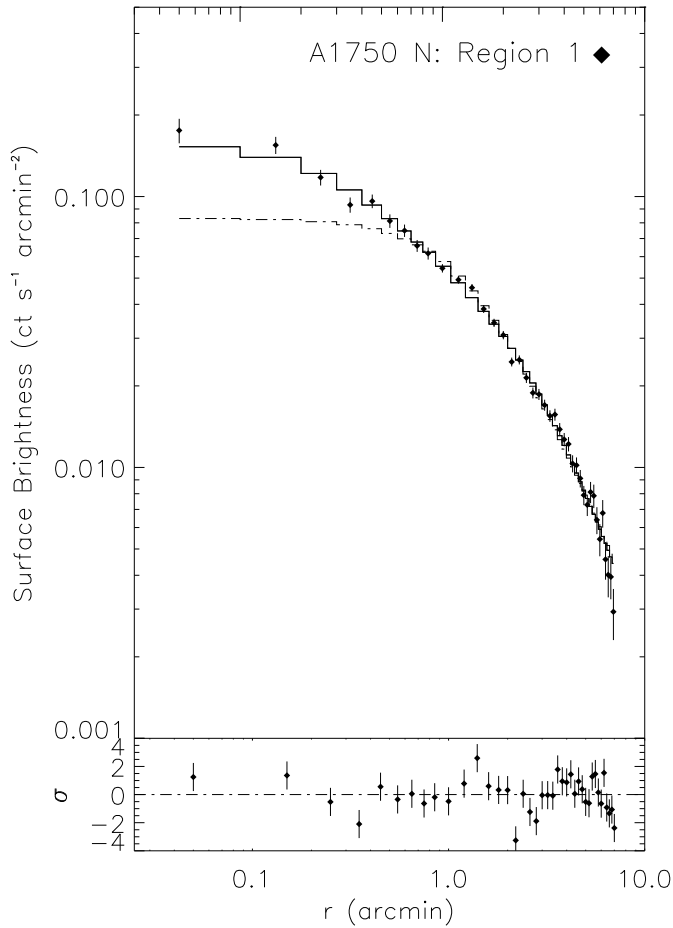}
\caption{The surface brightness profile of A1750 N, Region 1, binned as described in the text. It has been background subtracted and corrected for vignetting. The profile is shown with the best fit BB model; the residuals are shown in the bottom panel. The dash-dotted line shows the best-fit single \betamod\ to the outer regions.}
\label{fig:reg1sbprof}
\end{centering}
\end{figure}

\begin{figure}[!]
\begin{centering}
\includegraphics[scale=1.2,angle=0,keepaspectratio,width=\columnwidth]{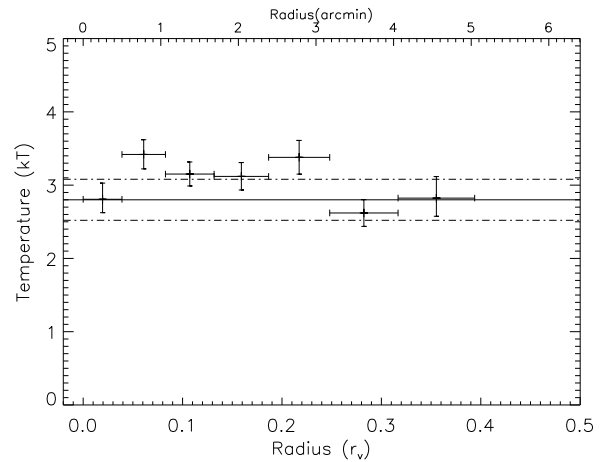}
\caption{ Temperature profile of A1750 N, Region 1. The straight line is the best fit global spectrum temperature, the dash-dotted lines denote $\pm 10\%$ of this value. Errors are at 1$\sigma$ for one interesting parameter. }
\label{fig:reg1tprof}
\end{centering}
\end{figure}


\subsection{A1750 N; R1}
\label{sec:reg1profs}
A single \betamod\ as described above, while being a relatively good description of the outer regions, is not a good description of the entire profile ($\chi^2 = 122.5$ for 37 degrees of freedom (d.o.f)). As is often found in these cases, progressively excluding the central regions improves the fit. The best fit \betamod, obtained by excluding the inner $\sim 1\arcmin$ yields $\chi^2 = 48.9$ for 27 d.o.f.; in this case $\beta=0.46$ and $r_c = 1\farcm4$.  We thus used the double isothermal BB model to fit the profile. This resulted in a much better fit to the {\it entire} radial range: $\chi^2 = 57.4$ for 35 d.o.f. The BB model fit is shown in Fig.~\ref{fig:reg1sbprof}; best- fit results are shown in Table~\ref{tab:betamodelfits}.

The best-fit outer $\beta$ value of the BB model fit is in excellent agreement with the 2D analysis in Sect.~\ref{sec:model2D} (results in Table~\ref{tab:2Dbetamod}), and with the ROSAT-derived results presented by Donnelly et al.~(\cite{donnelly01}). The fact that this is so, when we are not fitting exactly the same regions, is an indication that the results derived here are very robust.

The radial temperature profile is shown in Fig.~\ref{fig:reg1tprof}; note that this is the first such profile for this cluster. The cluster is detected with excellent $S/N$ up to $\sim 0.4 \rv$, the limit being due to the edge of the \xmm\  FoV. The temperature profile in this Region shows a surprising amount of variation. However, while the variations are significant in terms of their errors, they are actually quite small. For instance, the lowest temperature annular sector has $\kT = 2.6\ \keV$, and the highest $\kT = 3.3\ \keV$, an absolute variation of $0.7\ \keV$.

There is a dip towards the centre, which may be due to a cooling flow (discussed further below in Sect.~\ref{sec:a1750n}). From the second to the final annular sector, there is a smooth, gentle decline broken only by the fifth annular sector. However, the general features of this temperature profile are in good agreement with the spatially-resolved temperature map (Sect.~\ref{sec:tmap}).

\begin{figure*}[!]
\centering
\includegraphics[scale=1.0,angle=0,keepaspectratio]{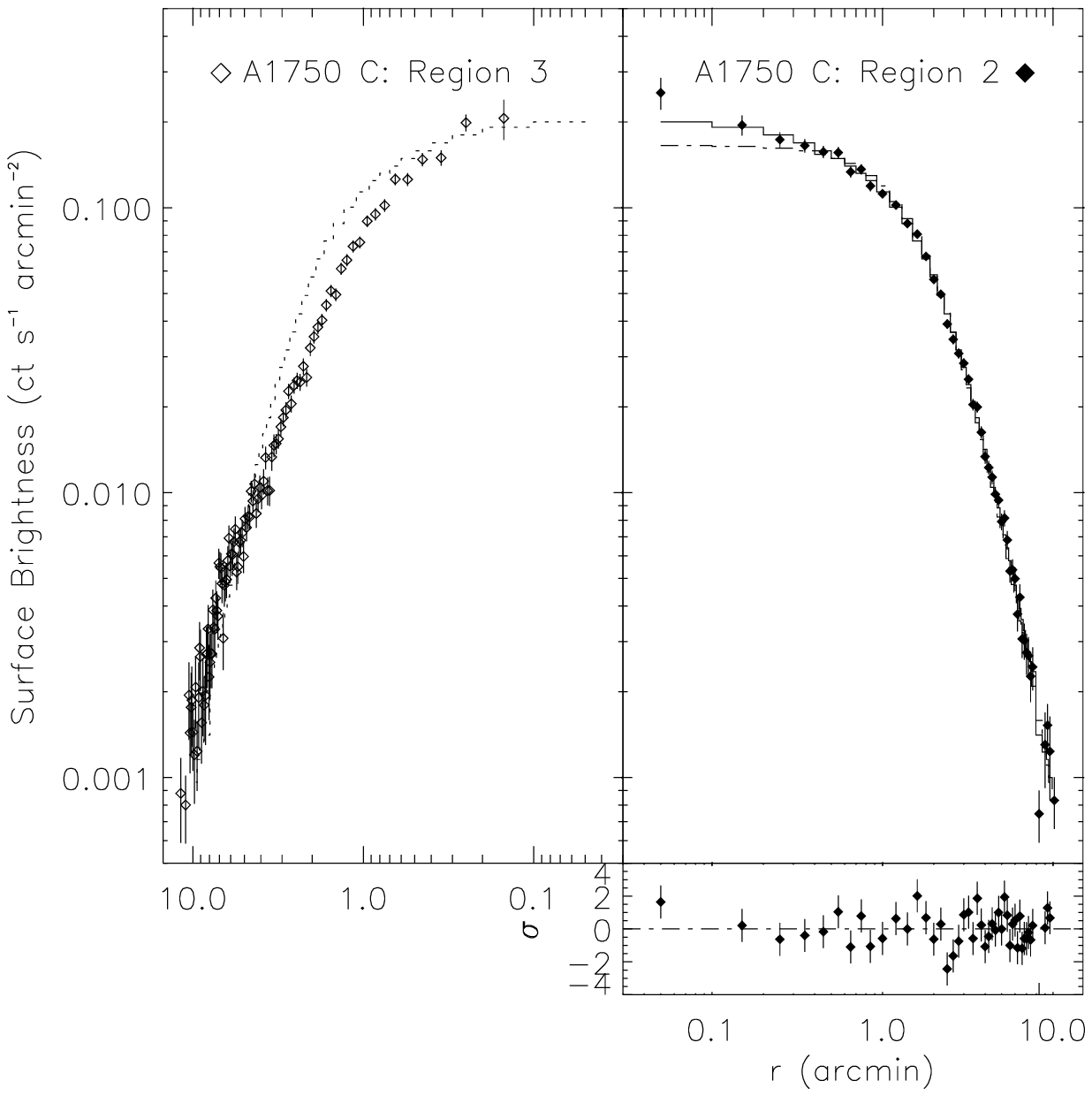}
\includegraphics[scale=1.0,angle=0,keepaspectratio]{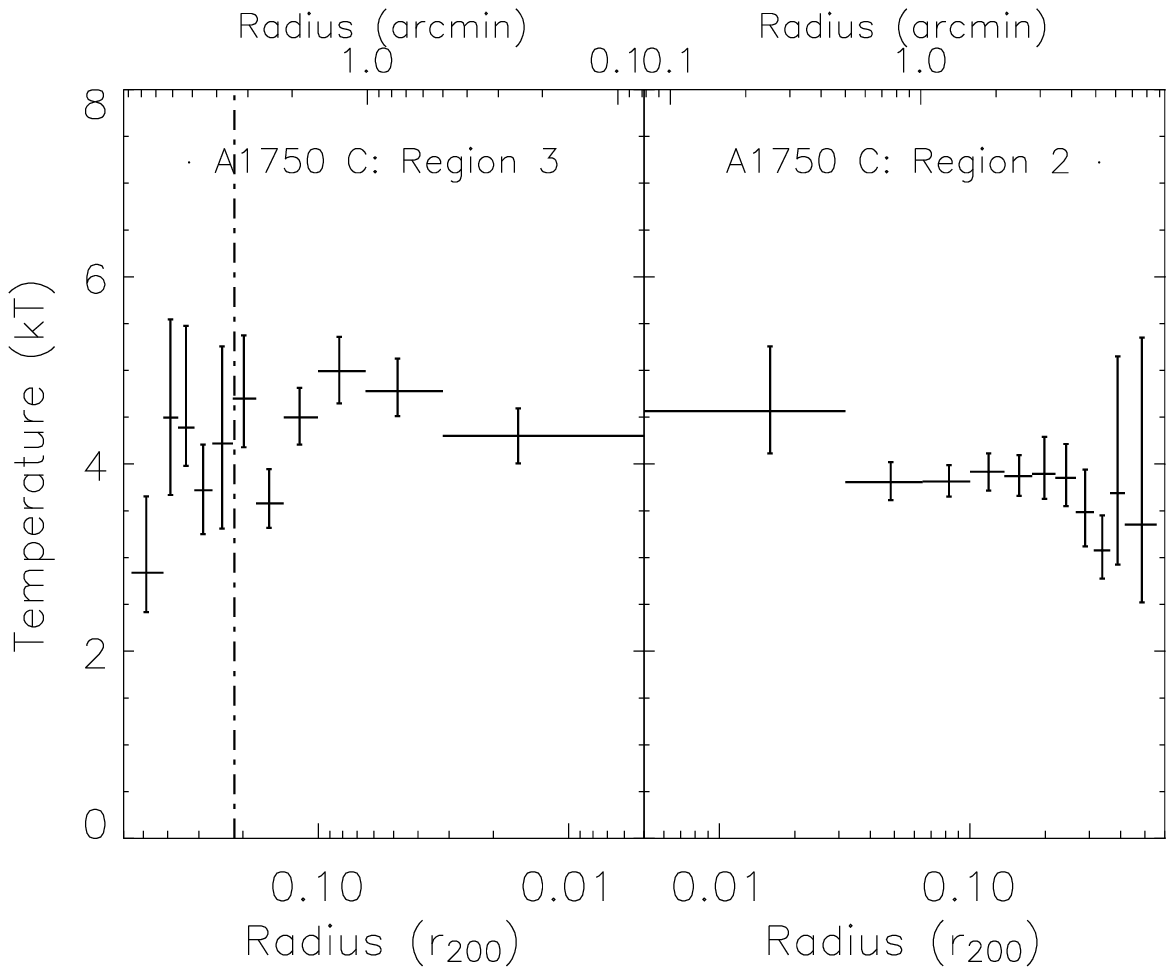}
\caption{The surface brightness (top) and temperature (bottom) profiles of Regions 2 and 3 of A1750 C. The right hand panel in each case shows R2, the left hand panel shows R3. For R2, the dot-dash line if the best-fit \betamod\ and the solid line is the best-fit BB model. The residuals are also shown. The dotted line overplotted on R3 is the best-fit BB model for R2.} 
\label{fig:a1750Cprofs}
\end{figure*}

\begin{table}
\begin{minipage}{\columnwidth}
\centering
\center
\caption{{\small Results of the BB model analytical fits to the gas surface
brightness profiles, errors are $90\%$ confidence.}}
\label{tab:betamodelfits}
\begin{tabular}{ l l l }
\hline
\hline

\multicolumn{1}{l}{ Parameter}  & \multicolumn{1}{l}{ R1} & \multicolumn{1}{l}{ R2 } \\ 
\multicolumn{1}{l}{ }  & \multicolumn{1}{l}{ A1750 N } & \multicolumn{1}{l}{ A1750 C W} \\ 

\hline

$n_{\rm H,0} ({\rm h_{50}^{1/2}~cm^{-3}})$ & $7.68 \times 10^{-3}$ & $5.89 \times 10^{-3}$ \\

$r_{\rm c}$ (\arcmin\ / kpc) & $1\farcm96_{-0\farcm17}^{+1\farcm83}$/ 254.4 & $2\farcm03_{-0\farcm12}^{+0\farcm15}$ / 263.5 \\

$\beta$    & $0.51_{-0.04}^{+0.05}$ & $0.69_{-0.02}^{+0.03}$    \\

$R_{\rm cut}$ (\arcmin\ / kpc) & $2\farcm04_{-0\farcm23}^{+2\farcm64}$ / 264.8 & $1\farcm05^{+0.21}_{-0.16}$ / 136.3\\

$r_{\rm c, in}$  (\arcmin\ / kpc) & $0\farcm12_{-0.04}{\footnote{The maximum value of $r_{\rm c, in}$ is fixed to $1\arcmin$.}}$ / 15.6 & $0\farcm10_{-0.02}^{~~~~~~a}$ / 13.0  \\

$\chi^2$/d.o.f & 57.4/35 & 42.2/42  \\

\hline
\end{tabular}
\smallskip

\end{minipage}
\end{table} %


\subsection{A1750 C; R2}

R2 extends to the eastern side  of A1750 C, and corresponds to the part of the cluster which shows the least amount of structure in isophotes and appears quite isothermal in the temperature map.

Once again, a single \betamod\ fails fully to describe the data ($\chi^2 = 56.7$ for 44 d.o.f.), although the central excess is far less evident than for R1. The best fitting single \betamod\ requires the exclusion only of the central 0\farcm7, and yields $\chi^2 = 36.6$ for 37 d.o.f. A BB model with parameters given in Table~\ref{tab:betamodelfits} is an excellent fit to the entire profile ($\chi^2 = 42.2/42$ d.o.f). The profile, together with the best fitting BB model, is shown in the right hand top panel of Fig.~\ref{fig:a1750Cprofs}.

The best-fit outer $\beta$ value of the BB model fit is in good agreement with the 2D analysis in Sect~\ref{sec:model2D}, and with the ROSAT-derived results presented by Donnelly et al.~(\cite{donnelly01}). Once again, this gives us good confidence in our results.

The temperature profile for R2 is shown in the right hand bottom panel of Fig.~\ref{fig:a1750Cprofs}; this is the first temperature profile for this cluster.

 The cluster is detected with excellent $S/N$ up to $\sim 0.6 \rv$. In this Region, as expected from the temperature map, the temperature profile is quite smooth, and shows little variability. 


\subsection{A1750 C; R3}
\label{sec:jump}

Here we come to the region of A1750 C which displays isophotal compression and, compared to R2, considerable temperature structure.

The surface brightness profile of R3 is shown compared to the profile from R2 in the left-hand top panel of Fig.~\ref{fig:a1750Cprofs}; in order to avoid smoothing of the discontinuity, we have not adopted logarithmic binning.. The difference between the profiles is striking: the R3 profile is more peaked, as is shown by the overplotted best-fitting BB model to R2. In addition, there appears to be a change of slope between 3 and 4 arc-minutes from the centre, the shape of which may indicate a discontinuity in the gas density profile. 

 For simplicity, we assume spherical symmetry. To quantify the discontinuity, we fitted the surface brightness profile of R3 with a radial density model composed of a $\beta$-model and a power law separated by a jump:
 
\begin{equation}\label{eq:jump}
 n (r) = \left\{ \begin{array}{llll}
		n_2 (r) & = & A_2 \left(1 + \frac{r^2}{r_c^2} \right)^{-3\beta/2} & \mbox{if $r<r_{cut}$} \\
\vspace{0.05cm}\\
		n_1 (r) & = & A_1 \left( \frac{r}{r_{\rm cut}}\right)^{\alpha}   &   \mbox{if $r>r_{cut}$} \\
		\end{array}
	\right.
\end{equation}

\noindent where 
\[
 n_1 |_{r_{\rm cut}}  =   X_{\rm jump} \times n_2 |_{r_{\rm cut}}.
\]

\begin{figure}
\begin{centering}
\includegraphics[scale=0.50,angle=0,keepaspectratio]{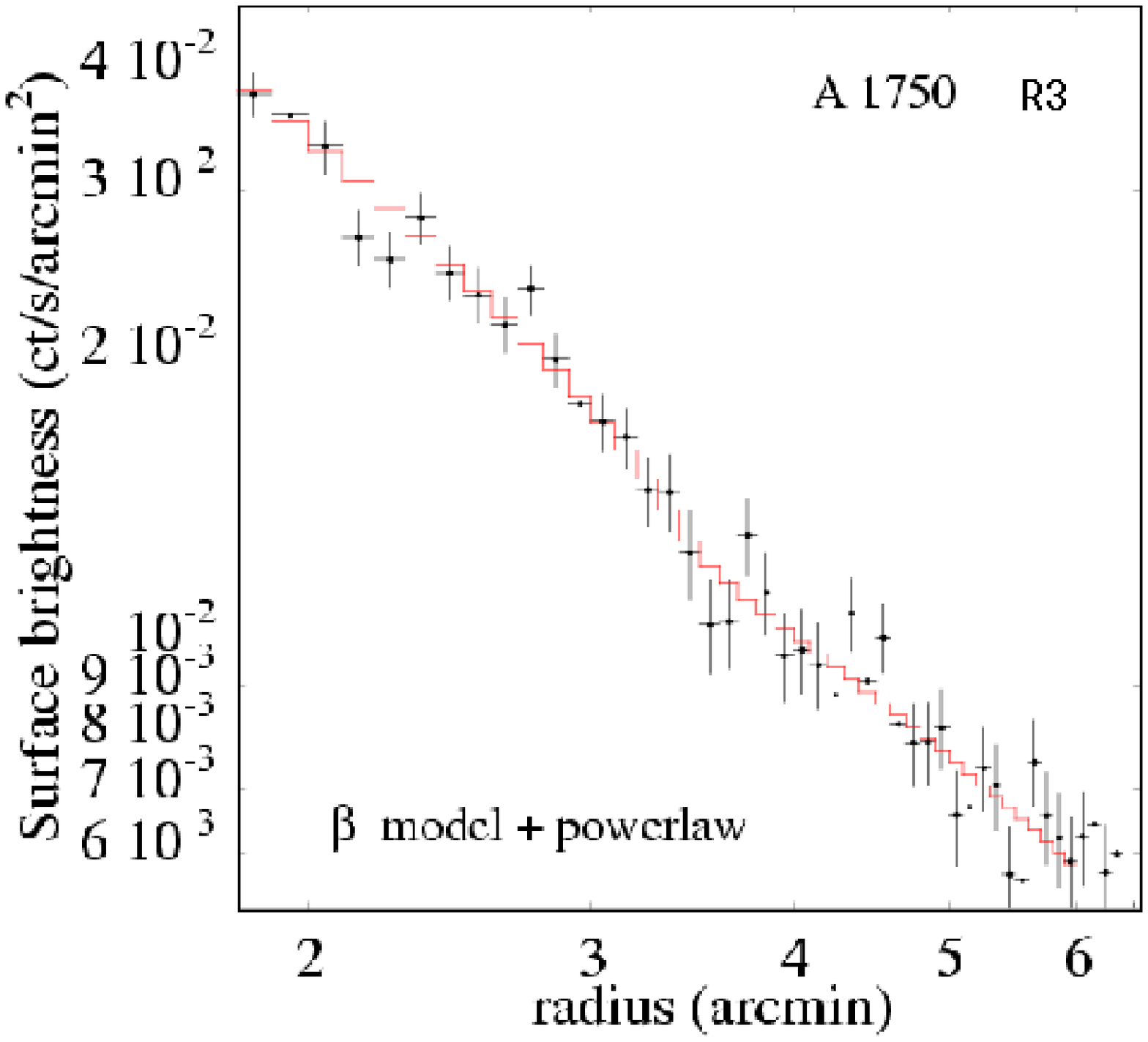}
\includegraphics[scale=0.55,keepaspectratio]{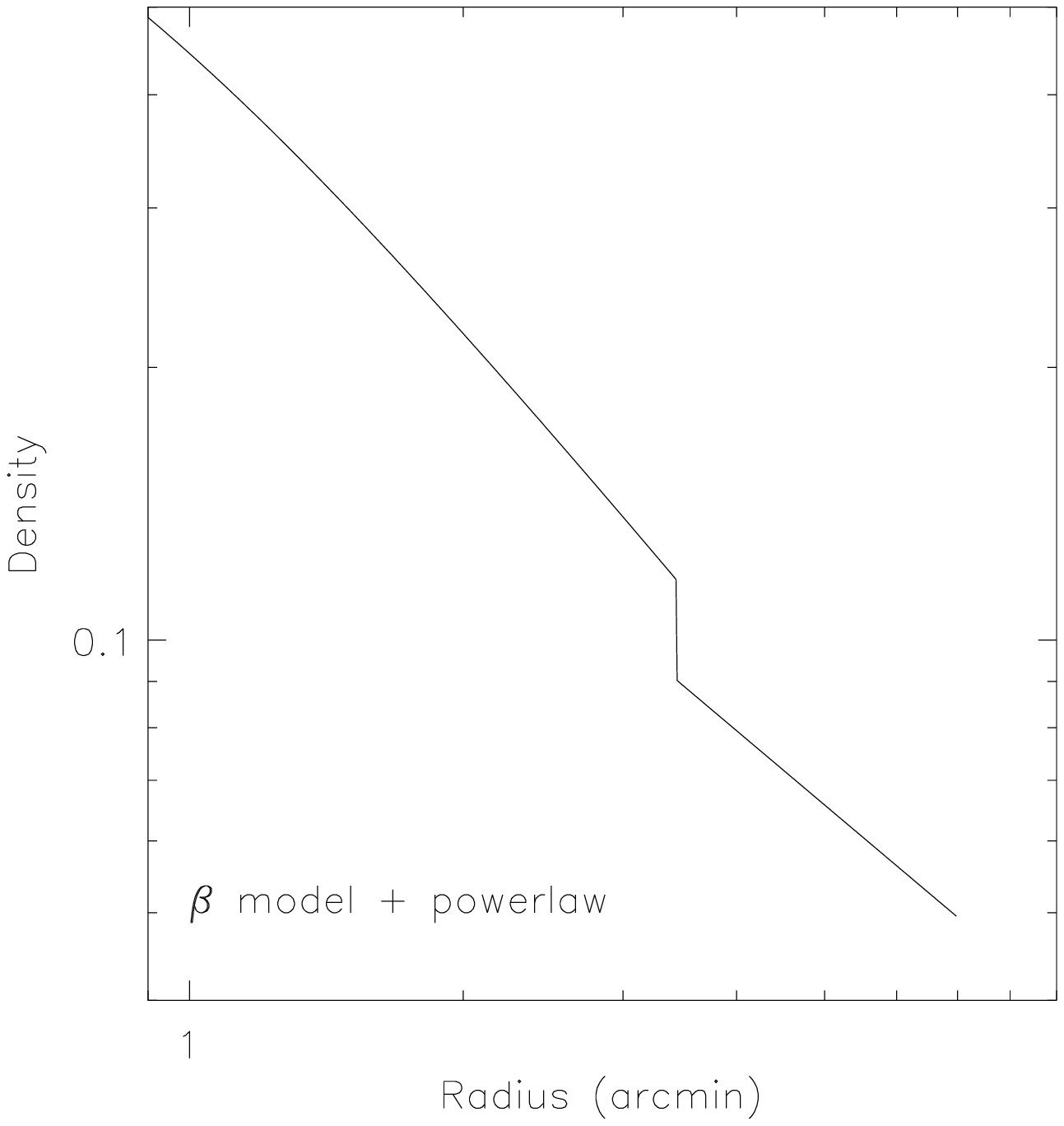}
\caption{{\bf Top:} Surface brightness profile of A1750 C, R3: data points and density model used to fit the data. {\bf Bottom:} Model used to fit the surface brightness profile (see Eqn.~\ref{eq:jump}). The model is normalised to 1 at the centre.}
\label{fig:sbprofA2}
\end{centering}
\end{figure}

In this model, $n_2$ and $n_1$ are the gas densities before and after the discontinuity, respectively. The free parameters of the fit were the slope $\beta$ and coreradius of the $\beta$-model, the slope $\alpha$ of the power law,  and the position ($r_{\rm cut}$) and amplitude $X_{\rm jump}$ of the density jump. The profile was fitted between bin 2 (10\arcsec) and bin 60 (6\arcmin). The best fit density model is shown both alone and superimposed on the data in Fig.~\ref{fig:sbprofA2}. The goodness of the fit gives an excellent  $\chi^2$= 51.4/53 d.o.f.  The best fit parameters are: $\beta$= 0.41, $r_c$=0\farcm6, $\alpha$=0.85. The jump position is $r_{\rm cut}$ = 3.44\arcmin\ and the amplitude of the density jump is $X_{\rm jump} = 0.78^{+0.09}_{-0.1}$, i.e., a $20\%$ jump in density. 

We also changed the bin range within which we fit the model and also used a second model composed of 2 power laws. In both cases we obtain consistent results within the errors, including jump amplitudes, indicating that the results are robust.

The orientation of the discontinuity  with respect to the line of sight is unknown: projection effects may act to reduce the apparent density jump at the discontinuity. If we do not see the discontinuity exactly perpendicular to the line of sight, the sharpness of the edge will be reduced. In other words, it is likely that our derived value for the density jump is a lower limit.


The radial temperature profile for R3 is shown in the bottom left panel of Fig.~\ref{fig:a1750Cprofs}. Here, the ring edges were chosen in order to match the density jump described above, and to have at least a 5$\sigma$ signal above the background in the [2.0 - 5.0] \keV\ band.

The temperature profile for this region is  quite variable. Even bearing 
in mind that both temperature profiles in Fig.~\ref{fig:a1750Cprofs} are 
plotted with a logarithmic x-axis, it is obvious that the profile of R3
is more variable than that of R2.
The  maximum of the temperature profile at $\sim 1\farcm5$  (in fact a 1$\sigma$ deviation) corresponds to the peak of the inner high-temperature arc seen in the temperature map (Figure \ref{fig:tmapwlet}). This temperature peak does not correspond to any significant structure in the surface brightness profile, even if the surface brightness profile at this distance from the centre does display some level of discontinuity. It is worth noting that within the errors we do not detect any temperature discontinuity at $3\arcmin.4$, the position of the jump in the density profile.  
We further note that there is a significant temperature jump between annuli 5 and 6. However this does not correspond to any significant jump in the observed surface brightness profile.


\subsection{Entropy profiles}

\begin{figure}[]
\begin{centering}
\includegraphics[scale=1.,angle=0,keepaspectratio,width=\columnwidth]{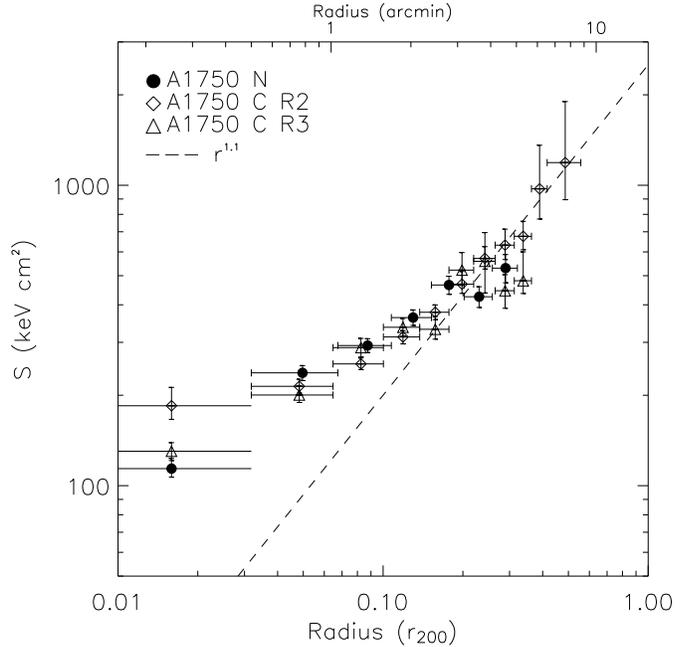}
\caption{The entropy profiles of Regions R1, R2 and R3. The dashed line corresponds to $S \propto r^{1.1}$, with an arbitrary normalisation. The profile for R3 has been limited to the range of our fit to the surface brightness ($r < 6\arcmin$).}
\label{fig:entropyprof}
\end{centering}
\end{figure}

With a view to making a comparison with relaxed clusters, we determined the entropy profiles of Regions R1, R2 and R3. The entropy was determined from the BB analytical model fit to each gas density profile and the observed (projected) temperature profiles, taking, as is now customary in this field, $S = T/n_e^{2/3}$. For R3, we use the gas density model described above in Sect.~\ref{sec:jump}, limiting the radial values to the fit limit of $r < 6\arcmin$. The resulting entropy profiles are shown, plotted in terms of the virial radius, in Fig.~\ref{fig:entropyprof}. Typical errors, corresponding to the error in each bin of the temperature profile, are indicated. The overplotted line shows the $S \propto r^{1.1}$ behaviour expected from shock heating (Tozzi \& Norman~\cite{tn01})

\section{Discussion}

\subsection{The current merging event}
\label{sec:current}

In this Section we discuss a possible scenario for the current merging event (i.e., that between A1750 N and A1750 C) based on simple assumptions. 

The weak enhancement in temperature between the two clusters (Fig.~\ref{fig:tmap}) is a strong indication that the clusters are just beginning to interact. 

We can attempt to calculate the Mach number and the collision velocity from the observed temperature variations. In this analysis, we follow the simple assumptions outlined in Markevitch et al.~(\cite{msv99}), viz: that the merger is symmetric in the plane of the sky, that the shocked gas is on average at rest with respect to the centre of mass, and that the shocked gas is nearly isothermal and in equipartition.  The Mach number and collision velocity can then be obtained by applying the Rankine-Hugoniot jump conditions for a one-dimensional shock, following Landau \& Lifshitz (\cite{LL59}; see also Markevitch et al. \cite{msv99}; Sarazin~\cite{sarazin02}):
\begin{equation}
\frac{1}{C} = \left[4 \left(\frac{T_2}{T_1} - 1\right)^2 +\frac{T_2}{T_1}\right]^{1/2} - 2 \left(\frac{T_2}{T_1} -1 \right) ; \\
\label{eq:machT}
\end{equation}
\begin{equation}
{\cal M}^2  = \frac{3C}{4 -C};
\label{eq:mach}
\end{equation}
\noindent  where Eqn.~\ref{eq:machT} corresponds to Equation 2 of Markevitch et al.~(\cite{msv99}) when $\gamma = 5/3$; $T_1$, $T_2$ are the pre-shock and post-shock temperatures and C = $n_2/n_1$ is the shock compression. 

For this rough estimate, we assume that the pre-shock temperature (k$T = 3.10 \pm 0.14\ \keV$) can be approximated by the average temperature on the sides of each cluster symmetrically opposite the shock region (see Markevitch et al.~\cite{msv99}), and the post-shock temperature (k$T = 5.12 \pm 0.73\ \keV$) is given by the observed global temperature of the middle region (Sect.~\ref{sec:globspec}).  We obtain C = 1.89 and a Mach number of ${\cal M} = 1.64$. 

Assuming that the surface brightness scales as the square of the density, using the above results we expect an increase of $\sim 3.6$ in the surface brightness between the clusters. This is not a huge effect. To compare directly with the observations we integrated the surface brightness perpendicular to the axis joining the cluster centres, and compared this to the same integration of an image of the best-fitting BB models. We then computed the ratio in brightness between the middle region and the region symmetrically opposite the clusters (in a similar fashion to the determination of the Mach number above), for both the image and the model\footnote{The definition of the outer regions is somewhat arbitrary since we do not know the real distance between the clusters}. The range of ratios we found (1.5 - 6.2) brackets the expected increase.

To first order, the time until first core passage ($t_{\rm merge}$) between A1750 C and A1750 N can be obtained by considering the distance of the two components and the distribution of the temperature and the density, even if projection effects are the major limitation in the accuracy of such an analysis. 

The centres of A1750 C and A1750 N lie at a projected distance of 675 $h_{50}^{-1}$  kpc. If we assume that the difference in radial velocity is exclusively due to the infall of A1750 N onto A1750 C (in the reference frame of A1750 C), the two systems would be so close --- well within the virial radius of each cluster --- that we should observe far larger density/surface brightness distortions and temperature enhancements in the middle region (as seen in e.g., Fig.~3 of Ricker \& Sarazin 2001). 
On the other hand, if we assume that the  physical distance is determined by the Hubble flow, meaning that the motion of the two clusters is exclusively radial, the difference in their mean radial velocities would yield a distance between the two clusters of $\sim 20$ $h_{50}^{-1}$ Mpc. This corresponds to nearly 10 times the virial radius of either cluster and thus the two should not display any sign of interaction whatsoever, in disagreement with the results described in detail above. The real distance surely lies in between these two values, and judging by the observed characteristics, is likely to be less than the virial radius of A1750 N ($\sim 1.7$ Mpc), using the  $r_{200}-T$ relation of Evrard et al.~(\cite{emn96} -- Eqn.~\ref{eq:emn96}).

Using Equation~1 of Markevitch et al.~(\cite{msv99}), we calculate a collision velocity of $1400$ km s$^{-1}$, which is roughly of the same order as the radial velocity difference between A1750 C and A1750 N. For a rough, order of magnitude estimate of $t_{\rm merge}$, we first assume the distance between the clusters to be the projected distance. In this case, we find $t_{\rm merge} = 0.47$ Gyr. If we now assume that the distance between the clusters is given by the estimated virial radius of A1750 N, we find $t_{\rm merge} = 1.3$ Gyr. 

We can also compare our results with simulations of merger events of systems of nearly equal mass (Roettiger et al. 1996,~\cite{roettiger97}; Ricker \& Sarazin~\cite{RS01}; Ritchie \& Thomas~\cite {RT02}; Teyssier 2002), which resemble very closely the observed characteristics. We thus expect the two clusters to lie at a real distance close to, but somewhat  lower than, their virial radii, around 1 $h_{50}^{-1}$ Mpc. This would imply that the clusters will reach core passage some time in the next $\sim 0.7$ Gyr.


\subsection{A1750 C: an unrelaxed cluster}

Four pieces of evidence lead us to conclude that A1750 C is not a relaxed cluster, and further, that the perturbed state is intrinsic to the cluster itself and is not connected to the nascent merger with A1750 N. These are:

\begin{enumerate}

\item {\bf The observed discontinuity in the gas density profile}. 

\noindent A shock front is the most natural interpretation for the density discontinuity observed in A1750 C, R3 (Sect.~\ref{sec:jump}). In addition, the entropy map and profile of R3 (Fig.~\ref{fig:entropyprof}) show exactly the type of behaviour expected from a shock (the denser side has higher entropy).

Heating shocks are expected on theoretical grounds when clusters collide. Furthermore, such shocks are expected to be relatively weak, with a distribution of Mach numbers peaked at ${\cal M}$ = 1.4 (Gabici \cite{gabici03}). Shocks were proposed to explain the peculiar temperature and density structures observed in Cygnus A and A3667 (Markevitch et al. 1999) with ASCA and ROSAT. A clear detection of a shock wave was obtained with Chandra in the 1E 0657-56 cluster (Markevitch et al. \cite{mark02}). Other weaker shocks were observed in A3667 (Vikhlinin, Markevitch \& Murray \cite{VMM01}) and A85 (Kempner, Sarazin \& Ricker \cite{ksr02}). 

By applying the Rankine-Hugoniot jump conditions (Eqn.~\ref{eq:mach})  we can obtain the Mach number of the shock in A1750 C with respect to the ambient gas (assuming again an adiabatic index for monoatomic gas 5/3). These conditions are valid if the gas within the shocked region is nearly isothermal and equipartition between electrons and ions applies.
From the density jump  in Sect.~\ref{sec:jump} we obtain ${\cal M}= 1.19^{+0.13}_{-0.09}$. The expected temperature ratio for a shock with such a Mach number is  $T_2/T_1 = 1.18$. Comparing the pre-shock  $T_1$=4.28$^{+0.71}_{-0.62}$ keV and  post-shock $T_2 = 4.59^{+0.56}_{-0.42}$ keV temperatures (see Sect. \ref{sec:jump}), we obtain a ratio $T_2/T_1 = 1.07(\pm0.21)$\footnote{Assuming symmetrical errors for T1 and T2}, and thus, within the  errors, the two independent measures are consistent. 

From the jump conditions we can also infer the pressure before and after the shock, obtaining a pressure jump of a factor 1.5. This demonstrates that the gas in cluster C is not in pressure balance and thus hydrostatic equilibrium conditions should not be applied.

The geometry of the shock suggests that it is related to a merging event that A1750 C has suffered in the past. The merger velocity is $v_{merge} \sim$ 500  km s$^{-1}$, and thus taking the distance from the centre, a rough estimate of the time of the A1750 C merger gives $t_{merge} \sim$ 0.9 Gyr. By comparing with simulations, the thermo-dynamical status this corresponds to an advanced phase of merger, at least 2 Gyr after core passage (Ricker \& Sarazin ~\cite{RS01}), which also corresponds to a sound crossing time\footnote{$t_{\rm s} = 6.6 10^8 ~(T/10^8 K)^{-1/2} (D/$Mpc) where $D$ is the cluster diameter  (see Sarazin \cite{sarazin} eq. 5.54)}. Given our approximations, possible projection effects and the physical changes in the medium during the merger, and the difficulty of comparing with simulations, we cannot really be more precise. 

 The observed temperature jump between bins 5 and 6 in the R3 temperature 
profile brings to mind a cold front. However, the propagation of a cold 
core into a hotter medium should also produce a density jump, as observed 
in e.g., A2142 (Markevitch et al.~\cite{mark00}). We have not been able to find
such a discontinuity  at this position, even by using profiles extracted in 
elliptical rings. However, given possible projection effects, we cannot 
entirely rule out the possibility of a cold front. Obviously, the better 
spatial resolution of a Chandra observation would enable us to provide a 
definitive answer to this question.

\item {\bf The elliptical isophotes at small scale, the offset of the core from the expected position given the larger scale structure, and the shift of the X-ray centroid from the position of the brightest cluster galaxy (BCG)} . 

These are particularly evident after subtraction of a 2D \betamod\ (Fig.~\ref{fig:resid2D}), and in Fig.~\ref{fig:fig1b}.

In their simulations of merging clusters Roettiger et al.~(1997) predict a late phase of the interaction, after core passage, when the dark matter and the gas separate and the gas sloshes in a potential which is starting to re-establish a new equilibrium configuration. This phase is long in the timeframe of the merger event (at least up to 5 Gyr in the Roettiger et al. simulations). This sloshing of the gas can produce an elongated distribution as observed in A1750 C as well as a centroid shift (see also Roettiger et al. 1996).  A similar gas distribution is observed in A1795 (Markevitch et al. 2001; Fabian et al. 2001; Ettori et al. 2002). In this case, however, the extent of the sloshing gas is 1/2 of that of A1750 C and is more closely connected with the central galaxy and the cooling flow. Note that Markevitch et al. (2001) and Ettori et al. (2002) concluded that the centre of A1795 is not in hydrostatic equilibrium even at larger scale. This is thus further strong observational evidence that A1750 C is in the late phase of an old merger event. 

\item {\bf The lack of evidence for a cooling flow}. 

\noindent The flat central temperature gradient (Fig.~\ref{fig:a1750Cprofs}) argues quite strongly against the presence of a cooling flow. We confirm this by computing the cooling time using Eqn.~\ref{eq:coolingtime}. This gives $t_{\rm cool} = 1.05 \times 10^{10}$ yr, of the same order as the Hubble time at the redshift of the cluster. Simulations of merging clusters suggest that only nearly head-on major mergers between objects of nearly the same mass can destroy cooling flows (e.g., Ritchie \& Thomas 2002). Thus the lack of cool gas in the centre of A1750 C is likely the result of an old -- and rather violent -- merger event.

\item {\bf The entropy profile, which appears to differ significantly from other high quality \xmm\  profiles of relaxed clusters.} 

\begin{figure}[]
\begin{centering}
\includegraphics[scale=1.,angle=0,keepaspectratio,width=\columnwidth]{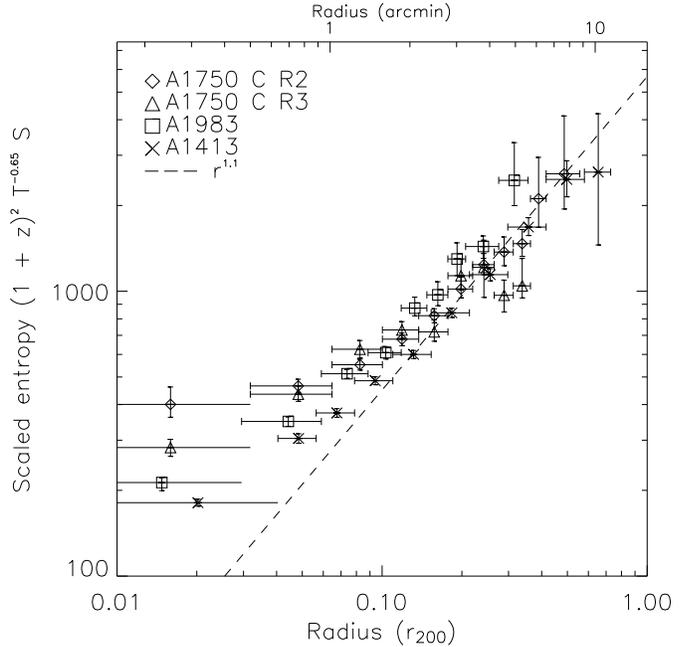}
\caption{The scaled entropy profiles of Regions R2 and R3 compared with the entropy profiles of A1413 and A1983 (Pratt \& Arnaud 2002, 2003). The profiles have been scaled with using the empirically-derived relation $S \propto T_X^{-0.65}$ of Ponman, Sanderson \& Finoguenov~(\cite{psf03}). The dashed line corresponds to $S \propto r^{1.1}$, with an arbitrary normalisation.}
\label{fig:entropyprof2}
\end{centering}
\end{figure}

We show in Fig.~\ref{fig:entropyprof2} the entropy profiles of R2 and R3, compared with the profiles of the relaxed clusters A1983 ($\kT = 2.1$ keV; Pratt \& Arnaud 2003) and A1413 ($\kT = 6.9$ keV; Pratt \& Arnaud 2002). The profiles have been scaled with the empirically-derived scaling relation ($S \propto T_X^{-0.65}$) of Ponman et al.~(\cite{psf03}). While both R2 and R3 appear approximately to converge to the $S \propto r^{1.1}$ value expected from shock heating (Tozzi \& Norman~\cite{tn01}) at a radius of $\sim 0.2 \rv$, each exhibits a very high scaled central entropy value for its temperature (or mass), compared to the scaled entropy of the relaxed clusters.

In form, the entropy profiles of R2 and R3 are qualitatively similar to those found in the merger simulations of Ricker \& Sarazin~(\cite{RS01}). The entropy map is also reminiscent of their simulations. The increase in core entropy due to a merger has also been seen in the simulations of Ritchie \& Thomas~(\cite{RT02}). The increase appears to be the result of the mixing of shocked, high entropy gas from the outer regions into the low entropy core of the cluster. Thus the entropy profile is another strong piece of evidence that A1750 C has not yet relaxed from a previous merger event.

\end{enumerate}

 Crucially, all of the evidence listed above solely concerns A1750 C. These data strongly suggest that the cluster is in the late phase of a merger for which the core passage happened 1-2 Gyr ago. The cluster is in the process of re-establishing equilibrium, but this is likely to be interrupted by the new merging event occurring with A1750 N.


\subsection{A1750 N}
\label{sec:a1750n}

The morphological analysis of A1750 N (Sect.~\ref{sec:model2D}) has shown  the elongated shape of the cluster, as well as an excess above a standard \betamod\ which appears considerably off-centre with respect to the overall cluster morphology.

In discussing the form of this cluster we should first mention that at
this off-axis angle PSF effects are expected to become important. In order
to have a qualitative idea of this effect, we have simulated the whole
observation in SCISIM, using as input the best-fitting 1D \betamod\ values
listed in Table~\ref{tab:betamodelfits}. The resulting image confirms that
emission from A1750 N should indeed be smeared by the PSF, preferentially
in the azimuthal direction.

With this in mind, we now turn to the ROSAT image of the cluster, shown in
Fig.~\ref{fig:rosatima}, and discussed in more detail below
(Sect.~\ref{sec:largescale}). It is evident that A1750 N seems elongated in
the NE direction; however, given the larger field of view of ROSAT, we
notice that there is another extended source just to the North of the
cluster. We can thus speculate that there is some contamination by
emission from this source, and indeed the 2D \betamod\ analysis of the
\xmm\ observation suggests extended residuals in the direction of this
source. Again we caution that PSF effects play a role at this distance
from the centre.

Finally, we cannot entirely rule out that A1750 N has been disturbed, or is
being disturbed, by interactions. It is possible that the observed
isophotal compression to the South of the centre (see figure \ref{fig:resid2D}) is due to the interaction with A1750 C (suggesting a non-zero impact parameter), but with the present observation we cannot definitively say whether there is interaction with the source to the North.

Donnelly et al.~(\cite{donnelly01}) speculate on the presence of a cooling flow in A1750 N. This is qualitatively in agreement with what we observe: a temperature drop is observed in the core of A1750 N (Fig.~\ref{fig:reg1tprof}), together with an excess above a standard \betamod.

In order to verify the presence of a possible cooling flow in A1750 N, we calculated the cooling time using (Sarazin 1986):
\begin{equation}
t_{cool} = 2.9 \times 10^{10} {\rm yr} \left(\frac{kT_X}{1~{\rm keV}}\right)^{1/2} \left(\frac{n_e}{10^{-3}}\right)^{-1}.
\label{eq:coolingtime}
\end{equation}
 
Using the central gas density obtained in Sect.~\ref{sec:reg1profs},  $n_e$(0)=$7.68 \times 10^{-3}$ cm$^{-3}$; Eqn.~\ref{eq:coolingtime} gives $t_{\rm cool} = 6.4 \times 10^9$ yr, about half of the age of the Universe at the redshift of the cluster. 
This is consistent with a weak cooling flow.
We thus fitted the MOS spectrum of the central bin with more complicated models:
\begin{itemize}
\item The sum of two MEKAL models absorbed by a column density fixed to the global spectrum fit value. The abundances of the two components are tied together.
\item The sum of a MEKAL model and a cooling flow model (MKCFLOW), with fixed absorption and abundances of the CF model tied to the thermal model. Also the upper temperature of the CF model is tied to the temperature of the MEKAL model.
\end{itemize}

The results are shown in Table~\ref{tab:otherfit} together with the simple MEKAL model results for this bin. 
There is not a significant probability that the 2T model or the MEKAL+CFLOW model are a better representation of the data. All fits are very good and the small mass deposition rate found by fitting the CF model is in agreement with a small cooling flow.

\begin{table}
\begin{minipage}{\columnwidth}
\centering
\center
\caption{{\small Multi-temperature and CF fits of the inner annulus of A1750 N. The $F$-test is computed against the fit for a single temperature absorbed MEKAL model. Errors are $90\%$ confidence.}}
\label{tab:otherfit}
\begin{tabular}{ l l l l}
\hline
\hline

\multicolumn{1}{l}{ Parameter}  & \multicolumn{1}{l}{ 1T} & \multicolumn{1}{l}{ 2T } & \multicolumn{1}{l}{ CF } \\

\hline

k$T_1$ (keV)&  $2.77_{-0.35}^{+0.45}$ & $3.01_{-0.41}^{+0.79}$ & $3.28_{-0.46}^{+1.4}$\\

k$T_2$ (keV)& ---& $0.73_{-0.23}^{+1.47}$ & ---\\

$Z/Z_{\odot}$ & 0.2$_{-0.16}^{+0.22}$ & 0.25$_{-0.15}^{+0.39}$ &  0.21$_{-0.14}^{+0.21}$ \\

$\dot{M}$ ($M_{odot}$ yr $^{-1}$) & --- & --- &  4.97 ($<9.6$) \\

$\chi^2$/d.o.f & 108.22/100 & 105.15/98 & 105.77/98\\
$F_{prob}$ & ---  & 75.6\% &  67.4\% \\

\hline
\end{tabular}

\smallskip
The low temperature of the CF model $T_{low}$ was 0.71 keV.
\end{minipage}
\end{table} %

Apart from the cooler core region, A1750 N displays a fairly uniform temperature distribution.  With the present data, all evidence suggests that the only physical process which is disturbing A1750 N today is the merger with A1750 C. However, we stress that a conclusive analysis needs an observation of the whole cluster, and the data presented here are limited by the off-axis position in the  \xmm\ field of view (as discussed further below in Sect.~\ref{sec:largescale}).


\subsection{At larger scale}
\label{sec:largescale}

\begin{figure}[b]
\begin{centering}
\caption{ROSAT/PSPC image smoothed with a Gaussian filter of $\sigma=20\arcsec$. The white circle traces the limit of the XMM field of view.}
\label{fig:rosatima}
\end{centering}
\end{figure}

The \xmm\ FoV has allowed us to observe the two main clusters of A1750 in one pointing. However, other structures have been observed at larger scale with {\em Einstein\/} (Jones \& Forman~\cite{jf99}). Figure \ref{fig:rosatima} shows a smoothed ROSAT/PSPC image obtained from the ROSAT archive: the superimposed circle indicates the \xmm\ FoV. To the south of A1750 C we observe another obviously extended source. This source has been observed in the optical and is discussed in Beers et al.~(\cite{beersetal91}), where it is called A1750 S; it lies at a projected distance of $\sim$ 2 Mpc from A1750 C and it is at the same redshift as A1750 C. Beers et al.~(\cite{beersetal91}), using optical virial mass estimators, suggest that A1750 S contributes $\sim 2\%$ to the total mass of the system. It is worth noting that the gas distribution to the south-western side of A1750 C appears elongated in the direction of A1750 S.

Fig.~\ref{fig:rosatima} also clearly indicates a further extended source just to the north of A1750 N, which is outside the FoV of our \xmm\ observation. We recall that the subtraction of a 2D \betamod\ (Sect.~\ref{sec:model2D}) suggested that there are further residual structures to the north of A1750 N, likely linked to this source. Other possibly extended sources are visible further to the north of A1750 N in the PSPC image. All these sources are aligned in the direction of the line joining the centres of the 3 main clusters, strongly suggesting a filamentary structure. Einasto et al. (\cite{einasto01}) found  that A1750 belongs to a rich super cluster composed of 7 members. We searched clusters in the redshift range 0.08-0.09 in a 16$^{\circ}\times16^{\circ}$ field centred on the centre of this super cluster and found 13 clusters, including A1750 which is located at the north-eastern edge.   

Given this observed large scale structure, it is not surprising that A1750 C  has suffered one (or several) previous mergers along the putative filament and indeed, statistical studies suggest a correlation between the level of substructures and the environment density (Schuecker et al. \cite{schuecker}). This adds a further piece of evidence in favour of our interpretation. This may well be the case for A1750 N as well, where its elongated gas distribution may also be the result of accretion along the filament.


\subsection{Caveat emptor}

The scenario outlined above explains in a self-consistent manner the data we have analysed. In particular, the observed temperature and surface brightness structure would be difficult to explain if the clusters have already past one another (as suggested for A3528 by Gastaldello et al.~\cite{gast}). We stress, however, that projection effects can be very important (e.g., as shown by Roettiger et al.~\cite{roettiger97}), and comparison with numerical simulations is necessarily a qualitative exercise. For a better understanding of this complex system, two more \xmm\ pointings, one each centred on A1750 N and A1750 S, would be very useful, as would deep optical observations.


\section{Summary and Conclusions}

We have reported the \xmm\  GT observation of the merging cluster A1750. The main conclusions of this work may be summarised as follows:

\begin{itemize}

\item In the morphological analysis with 2D $\beta$-models, we detect excess and off-centre emission in the cores of both A1750 N and C. We do not find any significant substructure in the region between the two clusters after subtraction of the $\beta$-models.  However, on close examination of the surface brightness in the region between the two clusters, we detect an increase of the same order as expected if the region between the clusters is a weak shock region (Sect.~\ref{sec:current}). We further detect excess residual emission to the north of A1750 N, which is likely (from comparison the \xmm\ and PSPC images) related to the larger scale structure of the system.

\item We have produced temperature maps by applying two different algorithms. We  measured a temperature increase of order $30\%$ in the region between A1750 N and C. A1750 N exhibits  a relatively smooth, uniform temperature distribution, but there are significant temperature variations within A1750 C, which appear intrinsic to A1750 C and not connected to the merger between the two clusters.

\item We use the 2D information to select interesting regions for a 1D analysis. For the regions corresponding to the relaxed, isothermal-looking parts of each cluster (R1 and R2), the 1D and 2D \betamod\ parameters are in excellent agreement with each other and with previous ROSAT analysis, although the \xmm\ data require an additional central component properly to fit the profiles. Both temperature profiles appear to decline modestly ($< 10\%$) with radius.

\item The density profile of the disturbed side of A1750 C (R3) is better described by a $\beta$-model and a power law, separated by a jump of order $20\%$. The temperature profile of this region is very irregular, but appears constant across the jump.

\item Compared to the scaled entropy profiles of relaxed clusters, the scaled entropy profiles of both regions of A1750 C exhibit a high scaled entropy value in the centre, similar to what is observed in merging cluster simulations. Only beyond $\sim 0.2 \rv$ does the profile of R2 attain the $S\propto r^{1.1}$ expected from shock heating.

\item In the assumption of an head-on merger between A1750 C and N we calculated the Mach number and velocity of the merger from the measured temperatures. With the resulting values (${\cal M}$ = 1.64 and $v_{merg} \sim$1400 km s$^{-1}$), we estimate that the two clusters will reach core passage in --- roughly --- less than 1 Gyr.

\item By applying the Rankine-Hugoniot jump condition to the measured density jump in A1750 C (R3) we confirm the detection of a weak shock of ${\cal M}$ = 1.2, in agreement with the measured temperature of the pre-shock and post-shock gas. Under simplified assumptions and by comparison with numerical simulations, we estimate that a merger --- intrinsic to A1750 C --- has occurred sometime in the past 1-2 Gyrs.

\item An inspection of the larger scale X-ray image suggests that A1750 N and C are the two main clusters within a filament which includes at least another small cluster at the same redshift of A1750 C (A1750 S). At still larger scale, A1750 lies in a filamentary supercluster containing, in total, 13 clusters.

\end{itemize}

We thus confirm that A1750 is a merging cluster in an early phase, when the two units have just started to interact and stand at a distance slightly lower than their virial radii. However, the global dynamical status of the system is far more complicated than expected. We conclude that the main cluster (A1750 C) has already suffered a previous merger and is now in the phase of re-establishing equilibrium. This phase will be interrupted by the current merger with A1750 N.  As far as we know, this is the first time such complex dynamical signatures have been observed in clusters at this merger phase.

The present day morphology of clusters may thus depend not only on on-going mergers (or the last merger) but also on the more ancient merging history, especially in dense environments. This has to be taken into account in the interpretation of statistical studies of cluster morphology.

\begin{acknowledgements}
We are very grateful to Dr. M. Arnaud for her participation in the scientific discussion and help in the analysis of the density profiles. We thank R. Teyssier for useful discussions. The authors thank the anonymous referee for the interesting comments which improved the manuscript.
This research has made use of the SIMBAD database,operated at CDS, Strasbourg, France. The paper is based on observations obtained with \xmm, an ESA science mission with instruments and contributions directly funded by ESA Member States and the USA (NASA). EPIC was developed by the EPIC Consortium led by the Principal Investigator, Dr. M. J. L. Turner. The consortium comprises the
following Institutes: University of Leicester, University of
Birmingham, (UK); CEA/Saclay, IAS Orsay, CESR Toulouse, (France);
IAAP Tuebingen, MPE Garching,(Germany); IFC Milan, ITESRE Bologna,
IAUP Palermo, Italy. EPIC is funded by: PPARC, CEA,CNES, DLR and ASI.
\end{acknowledgements}

\end{document}